\documentclass[reprint,aps,prl,superscriptaddress,amsmath,amssymb,floatfix,a4paper]{revtex4-1}
\usepackage{graphicx}
\usepackage{layout}
\usepackage{dcolumn}
\usepackage{braket}
\usepackage{bm}
\usepackage{color}
\usepackage{verbatim}
\usepackage{txfonts}
\usepackage{gensymb}

\usepackage{natbib}
\begin{document}

\title{In-Plane Magnetoconductance Mapping of InSb Quantum Wells }

\author{J. T. Mlack}
\affiliation{Center for Quantum Devices, Niels Bohr Institute, University of Copenhagen, and Microsoft Quantum Lab Copenhagen, 2100 Copenhagen, Denmark}
\affiliation{Department of Physics and Astronomy, Johns Hopkins University, Baltimore, Maryland 21218, USA}

\author{K.S. Wickramasinghe}
\affiliation{Homer L. Dodge Department of Physics and Astronomy, University of Oklahoma, 440 West Brooks, Norman, Oklahoma 73019-2061, USA}

\author{T.D. Mishima}
\affiliation{Homer L. Dodge Department of Physics and Astronomy, University of Oklahoma, 440 West Brooks, Norman, Oklahoma 73019-2061, USA}

\author{M. B. Santos}
\affiliation{Homer L. Dodge Department of Physics and Astronomy, University of Oklahoma, 440 West Brooks, Norman, Oklahoma 73019-2061, USA}

\author{C. M. Marcus}
\affiliation{Center for Quantum Devices, Niels Bohr Institute, University of Copenhagen, and Microsoft Quantum Lab Copenhagen, 2100 Copenhagen, Denmark}

\begin{abstract}
In-plane magnetoconductance of InSb quantum wells (QW) containing a two dimensional electron gas (2DEG) is presented. Using a vector magnet, we created a magnetoconductance map which shows the suppression of weak antilocalization (WAL) as a function of applied field. By fitting the in-plane field response of the 2DEG, we estimate material disorder and g-factor as a function of crystal direction. The in-plane WAL suppression is found to be dominated by the Zeeman effect and to show a small crystal-orientation-dependent anistropy in disorder and g-factor. These measurements show the utility of multi-directional measurement of magnetoconductance in analyzing material properties.
\end{abstract}


\maketitle
The widely studied III-V semiconductor indium antimonide (InSb) is an attractive material for a variety of electrooptical, spintronic, and topological applications, due its small bandgap, small electron effective mass, strong spin-orbit coupling (SOC), and large g-factor. The two-dimensional electron gas (2DEG) formed in a quantum well (QW) of InSb has been used as the basis of high-speed transistors \cite{AshleyProc,JMSOrr,AshleyElectronics} and spintronic devices \cite{HChen,Dedigama_2006} and can likely be used for topological qubits \cite{Lutchyn,Oreg}, possibly realizing some advantages over their InAs 2DEG counterparts \cite{Suominen, Nichele, Fornieri}.  

Electronic properties of InSb 2DEGs have been shown to be anisotropic with respect to crystal axes \cite{Jayathilaka,Kallaher, Chung, Ball, Mishima, AMGilbertson,F_Qu}, which has been attributed to preferentially oriented defects \cite{Chung,Ball, Mishima} and anisotropy of SOC \cite{Jayathilaka,Kallaher, AMGilbertson}. Anisotropy of the g-factor with crystal direction has also been demonstrated in bulk 3D InSb \cite{Barticevic,Bria,YFChen}, as well as other heterostructured materials \cite{Nefyodov}, but has not been previously reported in InSb 2DEGs to our knowledge.  

Relevant electronic properties can be extracted from quantum magnetotransport \cite{Kallaher,Dedigama_2007,Hikami,Iordanskii, Malshukov15,Malshukov16,Minkov,Cabanas, Studenikin,Minkov,Brinks,Yang,Ishida1,Ishida2,YuG,Dspirito}. Quantum interference between time-reversed backscattered paths of electrons in the presence of spin-orbit interaction leads to a positive contribution to the conductance in zero magnetic field, known as weak antilocalization (WAL) \cite{Hikami,Iordanskii}, which can be suppressed by a magnetic field applied either parallel or perpendicular to the 2DEG. These two field orientations affect WAL via different mechanism. A perpendicular field breaks time-reversal symmetry via orbital (Aharonov-Bohm) coupling to diffusive trajectories contributing to transport, while an in-plane breaks time-reversal symmetry predominantly via Zeeman coupling to spin \cite{Malshukov15,Malshukov16,Minkov,Marinescu_2017}, and depends on the g-factor. Finite thickness and interface roughness can also lead to orbital coupling from an in-plane field \cite{Malshukov15,Malshukov16,Minkov}. Measuring and fitting the magnetoconductance as a function of field direction allows one to obtain estimates of these quantities. Experimentally, the low field in-plane suppression has been observed in some III-V semiconductor 2DEG materials \cite{Cabanas,Studenikin,Minkov,Brinks,Meijer,Meng_2015,Takasuna_2017} and InSb thin films \cite{Yang,Ishida1,Ishida2}, but has not been reported in InSb 2DEG QWs, although some measurements have been done at higher fields \cite{Nedniyom,Litivenko,F_Qu}. 

In this Letter,  we report low-temperature magnetotransport in a 2DEG formed in an InSb quantum well [Fig.~1(a)] patterned into Hall bars [Fig.~1(b)] with different orientations on the wafer [Fig.~1(c)]. We investigate the effects of the strength and direction of an in-plane magnetic field on WAL. Mapping magnetoconductance as a function of in-plane field provides insight into anisotropy in the electronic properties with respect to crystal direction. Theoretical fits provide estimates of the spin orbit and phase coherence lengths, and fitting to in-plane field data estimates the disorder parameter and product of the g-factor times effective mass ratio.

\begin{figure}[h]
\centering
\includegraphics{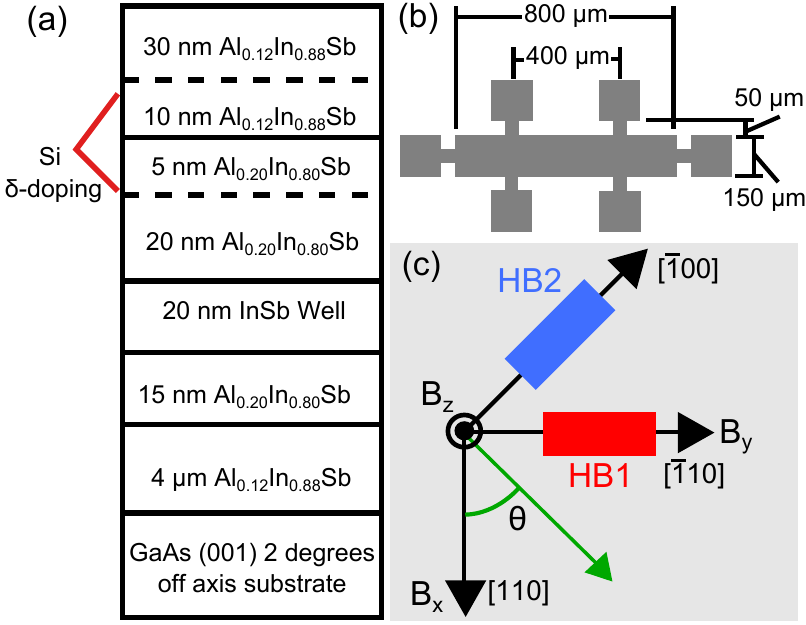}
\caption{\label{fig:1}
(a) Layer sequence of the InSb 2DEG quantum well. (b) Hall bar design with dimensions. (c) Geometric relationship of the fabricated Hall bars to the magnetic field axes of the vector magnet and the crystal structure of the InSb 2DEG. 
}
\end{figure}
Samples were measured using standard 4-wire low-frequency lock-in techniques in a dilution refrigerator with a 4-1-1 vector magnet system and a base temperature of 20 mK. The devices were fabricated from an asymmetrically doped InSb QW wafer, as shown in Fig.~1. Hall bars, with a width of $\rm{150\, \mu m}$ and voltage probes separated by $\rm{400\,\mu m}$, were defined using a ${\rm H_{2}O_{2}:C_{6}H_{8}O_{7}:P_{3}PO_{4}}$ (0.07:0.87:0.39) wet etch. The design is depicted in Fig.~1(b). Ohmic contact was made by electron-beam evaporation of Ti/Au. Classical Hall measurement yielded electron density $\rm{2.21 \times 10^{11}  cm^{-2}}$ and mobility 154,000  $\rm{cm^{2}/V s} $.  One of the Hall bars, HB1, was oriented along the $[\bar{1}10]$ crystal axis, while the other, HB2, was oriented along the $[\bar{1}00]$ axis.

We first investigate magnetoconductance in a perpendicular field with zero in-plane field (Fig.~2). A peak in conductance around zero perpendicular field was found for both Hall bar orientations, with a turnover at $\sim$~0.5~mT from decreasing to increasing conductance with field. This minimum in conductance corresponds to a crossover from WAL to weak localization (WL) corrections. We use the Rashba-dominated Iordanskii, Lyanda-Geller, Pikus (ILP) model \cite{Iordanskii} to estimate phase-coherence length, $l_{\phi}$, and spin-orbit length, $l_{so}$. Values of these parameters, listed in Table I, are important for the analysis of the in-plane field data and can also be used to calculate the spin-orbit energy $\Delta_{so} = \hbar^2 k_{f} \diagup m^{*} l_{so}$\cite{Iordanskii,Cabanas,Dspirito}, where $k_{f}$ is the two-dimensional (2D) Fermi wave vector.

\begin{figure}[h]
\centering
\includegraphics[scale=1]{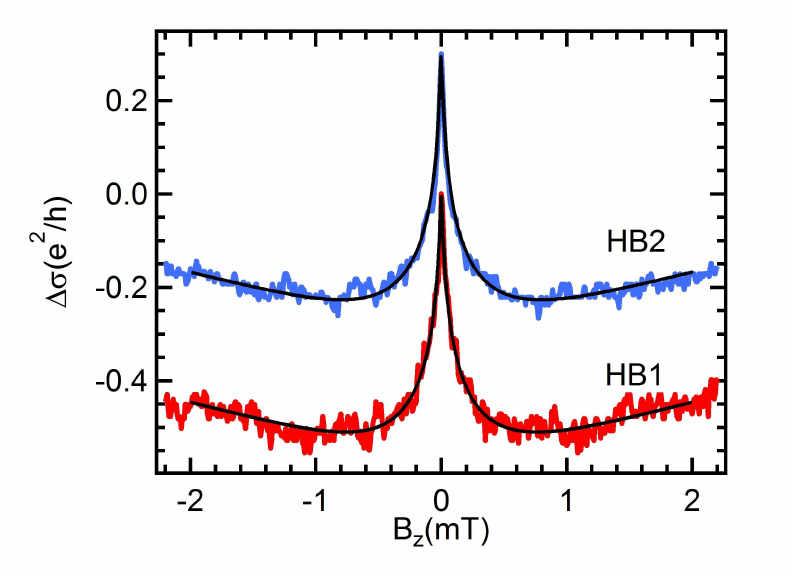}
\caption{\label{fig:2}
Measurement of the conductance correction in a perpendicular field resulting from weak antilocalization and weak localization, fit with the Rashba dominated ILP model (black)\cite{Iordanskii} for both HB1 (red) and HB2 (blue and offset by +0.3 $e^{2}/h$).
}
\end{figure}

\begin{table}[h]
\begin{tabular}{ c c c c c }

\hline
\hline

Sample & \hspace{6pt} $l_{e} (\mu m)$ \hspace{6pt} & \hspace{6pt} $l_{\phi} (\mu m)$ \hspace{6pt} & \hspace{6pt}$l_{so} (nm)$ \hspace{6pt}   & \hspace{6pt} $\Delta_{so} (meV)$ \hspace{6pt}  \\
\hline

HB1 & 1.22 & 9.12 & 976 & 0.667 \\
\hline
HB2 & 1.22 & 9.46 & 960 & 0.678 \\
\hline
\hline

\end{tabular}
\caption{Mean free path obtained from hall measurement and WAL ILP fit parameters from the fits displayed in Fig.~2}
\label{tab:1}
\end{table}

\begin{figure*}[ht]
\centering
\includegraphics[scale=1]{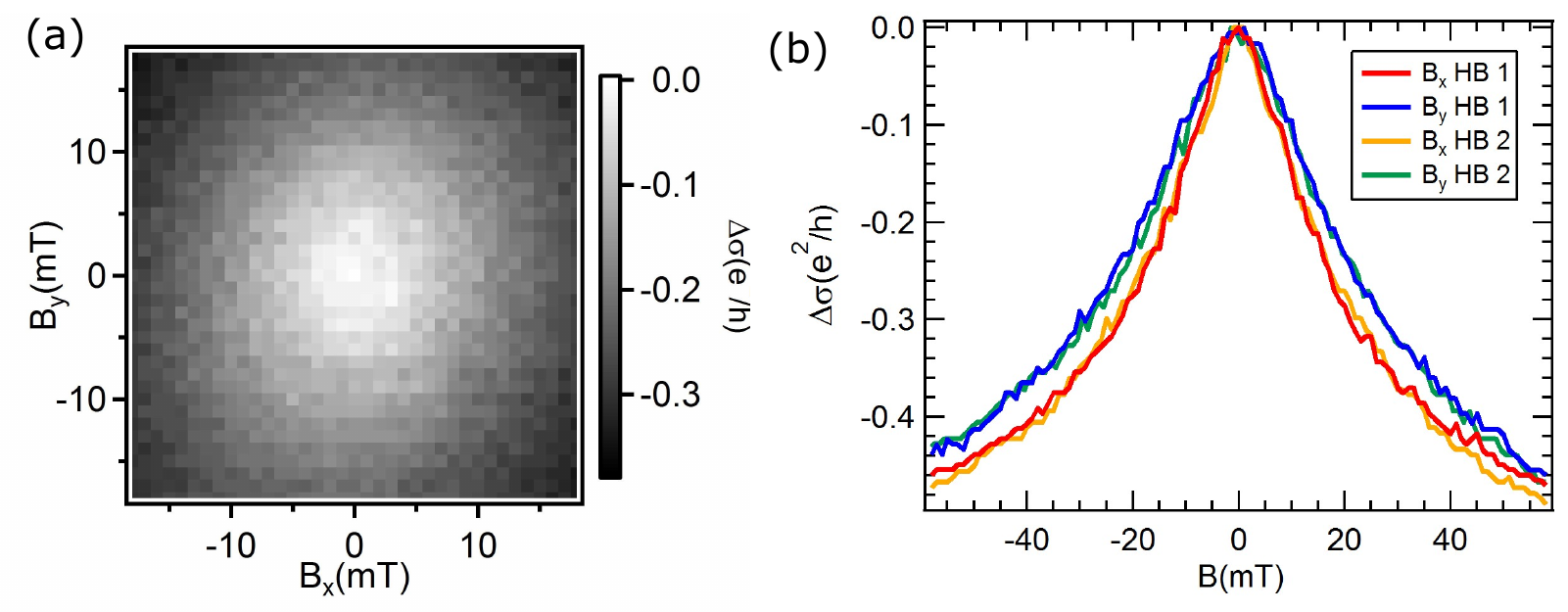}
\caption{\label{fig:3}
(a) Map of experimental measurement of the quantum correction to conductivity as a function of in-plane magnetic field for HB1. For each point in this map $B_{x}$ and $B_{y}$ were set and then $B_{z}$ was swept. The values plotted in the map are associated with conductance maxima of the $B_{z}$ sweep over a small range at each point, ensuring that the observed conductance suppression is only due to the effects of an in-plane field. (b) Comparison of the 0\degree and 90\degree axial in plane magnetoconductance, $[110]$ and $[\bar{1}10]$ crystal axes respectively, for both HB1(blue and red) and HB2(green and orange).
}
\end{figure*}

To study the effects of applying an in-plane field, maps of the magnetoconductance response were measured as a function of in-plane field, shown in Fig.~3a for HB1. For each point on the map, $B_{z}$ was independently swept in the small range, creating a three dimensional field data set that was then reduced to two dimensions. The map is predominantly symmetric in field directions $B_{x}$ and $B_{y}$. Any deviation from circular is difficult to resolve in this plot. Detecting asymmetry requires a more careful means of measurement. To do so, we examine magnetoresistance for HB1 and HB2 purely along $B_{x}$ and $B_{y}$ directions, corresponding to crystal directions $[\bar{1}10]$ and $[110]$, respectively, as shown in Fig. 3(b). For both devices the traces along the $[\bar{1}10]$, blue for HB1 and green for HB2, are in close agreement. Similarly the data along the  $[110]$ axis, red for HB1 and orange for HB2, are in close agreement. However, the data along each axis clearly shows a different magnetoconductance response. This indicates a response that is independent of the transport axis, but is tied to the crystal axis.

Using the theory of Minkov {\it et al.}~\cite{Minkov}, it is possible to fit these in-plane field traces and quantitatively study this anisotropy. Specifically it is possible to extract a structures's g-factor as well as an interface-roughness parameter, $\gamma_D$, reflecting random, zero-average orbital coupling due to a purely in-plane magnetic field. The expression used in the following analysis is a modified version of the equation developed by Caba\~nas {\it et al.}~\cite{Cabanas}, which assumes a known effective mass, but considers only a purely in-plane field, $B_{\perp} = 0$, yielding

\begin{align}
\label{eq:bpara}
\ \Delta\sigma=\sigma(0,B_{\parallel}) - \sigma(0,0) &= \frac{e^2}{ 4 \pi^2 \hbar }\left.\Bigg[ 2 \text{ln} \left( \frac{ B_{\phi,t} + B_{so} }{ B_{\phi} + B_{so} } \right)\right. \nonumber \\
&\left.+ \text{ln} \left( \frac{ B_{\phi,t} + 2B_{so} }{ B_{\phi} + 2B_{so} } \right)- \text{ln} \left( \frac{ B_{\phi,s} }{ B_{\phi} } \right)\right. \nonumber \\
 &\left.+S(B_{\phi,t}/B_{so})-S(B_{\phi}/B_{so})  \right. \Bigg]
\end{align}
The parameters $B_{\phi}$ and $B_{so}$ are the phase coherence and spin-orbit breaking fields, which are calculated from $l_{\phi}$ and $l_{so}$\cite{Cabanas} as
\begin{equation}
\label{eq:B_phi}
B_{\phi} = \frac{\hbar}{4 e l_{\phi}^2}
\end{equation}
and
\begin{equation}
\label{eq:B_so}
B_{so} = \frac{\hbar}{4 e l_{so}^2}
\end{equation}
 All in-plane fits were performed using the $l_{\phi}$ and $l_{so}$ obtained for the individual Hall bar. The term $S(x)$ is defined as:

\begin{equation}
\label{eq:s_term}
S(x) = \frac{8}{\sqrt{7+16x}}\left[\text{arctan} \left(\frac{\sqrt{7+16x}}{1-2x}\right)-\pi\Theta(1-2x) \right]
\end{equation}

The terms $B_{\phi,t}$ and $B_{\phi,s}$ represent the characteristic breaking fields of the triplet and singlet channel contributions to the conductivity and are defined as
\begin{equation}
\label{eq:bphit}
B_{\phi,t} = B_{\phi} + \gamma_D B_{\parallel}^2
\end{equation}
and
\begin{equation}
\label{eq:bphis}
B_{\phi,s} = B_{\phi,t} + \frac{1}{4 \hbar B_{so}} \left(\frac{m^{*} g \mu_{B} B_{\parallel}}{e l_{e} k_{f}} \right)^2,
\end{equation}
where $l_{e}$ is the mean free path and $\mu_{B}$ is the Bohr magneton. The two free parameters in the fit are $\gamma_D$ and $g m^{*}/m_{e}$, the Zeeman term. We note that the form of the fitting function is proportional to the square of  $g m^{*}/m_{e}$, making the sign of the g-factor inaccessible. 

For analyzing in-plane field effects, it is important to remove effects of small (sub-degree) misalignment of the sample plane with the magnetic field axes. To reduce any offsets, including possible field-dependent misalignment, $B_{z}$ was swept from $\pm$1mT for every point of in-plane measurement in order to measure the conductance peak associated with WAL. For each in-plane point the conductance value of the $B_{z}$ peak maximum was taken as the appropriate in-plane value. 

\begin{figure}[h]
\centering
\includegraphics[scale=1]{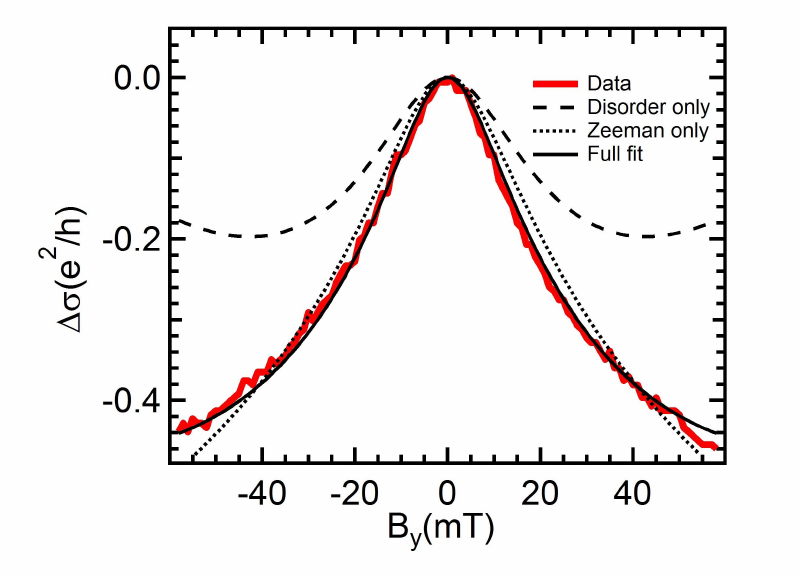}

\caption{\label{fig:4}
Fits (black), performed using Eq.~(1), of in-plane magnetocondutance data (red) taken on HB1 with the field oriented parallel to the Hall bar. The three fits represent fitting with both Zeeman and the disorder parameter (solid line), just the disorder parameter (dashed line), and just the Zeeman term (dotted line). 
}
\end{figure}

Corrected in-plane data (red), from HB1 with the field sweeping in-plane and parallel to the Hall bar, and fits (black), using Eq.~1, are shown in Fig. 4. The location of the turnover from WAL to WL in the data emphasizes the importance of correcting for offsets in measuring in-plane magnetoconductance. At 0.5~mT in Fig.~2, WAL has been completely suppressed, but in Fig.~4, even at 20mT, WAL has not been fully suppressed.

The fits shown in Fig.~4 were performed including both disorder and Zeeman terms (solid), only the disorder term (dashed), and only the Zeeman term (dotted). While a comparison of the three fits indicates that the Zeeman term dominates the effect, it is clear that disorder must also be taken into account to fully account for the behavior.

The best fit obtained that includes both parameters yields $g m^{*}/m_{e}$ = 0.91 and $\gamma_{d}$ = 0.0029$\rm{T^{-1}}$.  In Nedniyom {\it et al.}~\cite{Nedniyom}, a calculation is performed to estimate the effective mass ratio of electrons in a Rashba dominated InSb 2DEG. Extrapolating from these calculations, we find $m^{*}/m_{e} \approx 0.02$. Using this value of $m^{*}/m_{e}$ and the fitting parameter $gm^{*}/m_{e}$, the g-factor for in-plane parallel to the sample is $\vert g \vert= 45$. Although, the disorder term provides a smaller contribution to the fit, it is still a useful tool in analyzing magnetoconductance suppression as a function of crystal direction. 

\begin{figure*}[ht]
\centering
\includegraphics{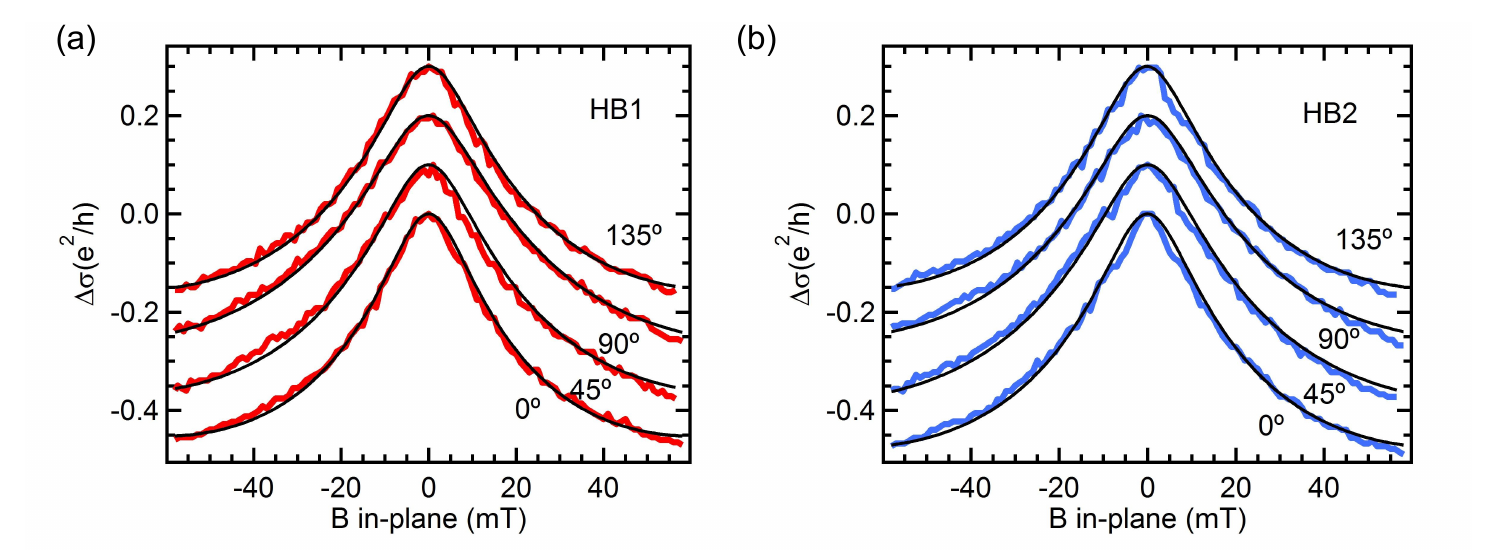}

\caption{\label{fig:5}
Data and associated fits, using Eq.~(1), for in-plane field measurements on both Hall bars. Each trace is in the plane of the 2DEG at a different angle counter-clockwise from the $B_{x}$-axis. Fit values are listed in Table II. (a) In plane field measurements and model fits of HB1, oriented along the $[\bar{1}10]$ crystal axis and parallel to the $B_{y}$(90\degree) axis. 
(b) In plane field measurements and  model fits of HB2, oriented along the $[\bar{1}00]$  crystal axis and parallel to 135\degree counter-clockwise from the $B_{x}$-axis.
}
\end{figure*}

In order to determine the properties as a function of crystal direction, measurements and fitting of in-plane field data were performed on HB1 and HB2 at  0\degree, 45\degree, 90\degree and 135\degree from the [110] axis  in the [001] plane (Fig.~5). The traces, and their respective fit curves, are shifted upwards for each increasing angle, starting at 0\degree, by +0.1 $\rm{e^{2}/h}$. The WAL suppression in both Hall bars starts to saturate at 50 mT and the data qualitatively show a similar field response between both Hall bars. 

\begin{table}[ht]
\begin{tabular}{ c c c c c }

\hline
\hline

Sample & \hspace{6pt} Angle \hspace{6pt} & \hspace{6pt} Crystal Axis \hspace{6pt} & \hspace{12pt} $g m^{*}/m_{e}$ \hspace{12pt}   & $\gamma_{d} (T^{-1})$ \\
\hline

HB1 &0\degree &$[110]$ & 1.15 & 0.0051 \\
HB1 &45\degree &$[010]$ & 1.02 & 0.0036 \\
HB1 &90\degree & $[\bar{1}10]$ & 0.91 & 0.0029 \\
HB1 &135\degree & $[\bar{1}00]$ & 1.03 & 0.0038 \\
\hline
HB2 &0\degree  & $[110]$ & 1.10 & 0.0042 \\
HB2 &45\degree & $[010]$& 0.96 & 0.0031 \\
HB2 &90\degree &$[\bar{1}10]$ & 0.90 & 0.0034 \\
HB2 &135\degree &$[\bar{1}00]$ & 1.00 & 0.0042 \\
\hline
\hline

\end{tabular}
\caption{Fit parameters obtained from fitting data presented in Figure 5 using Eqn. (1) and values for $l_{\phi}$ and $l_{so}$ shown in Table I}
\label{tab:2}
\end{table}

Fit parameters from Fig.~5 are displayed in Table II, and quantitatively show the angular dependence of the data. The fit values for $g m^{*}/m_{e}$ are each within $\pm$10 percent of the average value, 1.009. Using the aforementioned value for $m^{*}/m_{e}$, the average g-factor magnitude is 48, consistent with values in the literature\cite{Nedniyom,Barticevic,Bria,YFChen,F_Qu}. The fit values show a small sinusoidal anisotropy, with a maximum $g m^{*}/m_{e}$ in the $[110]$ direction and minimum in the $[\bar{1}10]$ direction. This behavior is comparable to the data from an asymmetrically doped GaAs QW in Nefyodov et al. \cite{Nefyodov}. In the article, they attribute the anisotropy to the electric potential created from the single sided doping, which is also the case with the QW presented here. It should be emphasized that the anisotropy is the same for both HB1 and HB2, thus we conclude that the observable anisotropy follows the crystal axis, not the Hall bar axis. 

Table II reveals that values for $\gamma_{d}$ show a much larger deviation, namely $\pm$20 percent of the average, $\gamma_{d}$ = 0.0038 $\rm{T^{-1}}$. A possible explanation for these large fluctuations is known directional dependence of mobility, being higher along the $[\bar{1}10]$ axis than the $[110]$, as discussed in Refs.~\cite{Ball} and \cite{Mishima}. Atomic force microscopy of the 2DEG material did not yield any obvious preferential defects. However, previous work \cite{Ball,Mishima} shows that the defects are buried in the material and best observed using a transmission electron microscope.

In conclusion, the in-plane magnetotransport of InSb QW Hall bars has been measured. The magnetoconductance response is qualitatively and quantitately anisotropic in crystal/applied field axis and independent of transport axis. Using WAL models \cite{Iordanskii,Cabanas}, estimates of the spin-orbit length, phase coherence length, g-factor, and disorder parameter of the 2DEGs were obtained. The large g-factor estimates agree with the known value for InSb and show a small anisotropy in the plane of the 2DEG that is strictly crystal dependent. The values obtained for the disorder parameter show a similar crystal dependence. In addition, these results show that the study of WAL can be used as a novel method to understand the material properties of 2DEG structures. These measurements serve as a first step in quantifying the crystal dependence of the electronic properties of InSb QWs and provide insight on how to orient devices for future measurements.


We thank Ray Kallaher, Morten Kjaergaard, Henri Suominen, Marina Hesselberg, and Shivendra Upadhyay for valuable conversations on measurement and fabrication. This work was supported under the National Science Foundation Grant No. DGE-1232825 (J.T.M.). The research was supported by Microsoft, the Danish National Research Foundation and the European Commission. C.M.M.~acknowledges support from the Villum Foundation. We would also like to thank 
\bibliographystyle{unsrt} 
\bibliography{SO_Paper_Bibtex}
\nocite{Golub,Glazov_Golub,Uddin}

\end{document}